\documentstyle[aps,epsfig,prl,floats]{revtex}
\begin{document}
\title{Strong-Coupling Expansions for Multiparticle Excitations:\\ 
Continuum and Bound States}
\author{
        Simon Trebst$^{(a)}$\cite{Bonn}, 
        Hartmut Monien$^{(b)}$\cite{Bonn}, 
        Chris J.~Hamer$^{(c)}$\cite{cjh},
        Zheng Weihong$^{(c)}$\cite{zwh}, and 
        Rajiv R.~P.~Singh$^{(d)}$ 
}
\address{
$^{(a)}$ Bell Labs, Lucent Technologies, Murray Hill, NJ 07974\\
$^{(b)}$ Yukawa Institute for Theoretical Physics, Kyoto University,
Kyoto, 606-8502\\
$^{(c)}$ School of Physics, University of New South Wales, Sydney NSW
2052, Australia\\
$^{(d)}$ Department of Physics, University of California, Davis, CA
95616\\
}
\twocolumn[\hsize\textwidth\columnwidth\hsize\csname
@twocolumnfalse\endcsname
\date{\today}
\maketitle
\widetext

\begin{abstract}
We present a new linked cluster expansion for calculating
properties of multiparticle excitation spectra to high orders.
We use it to obtain the two-particle spectra for systems of coupled
spin-half dimers. We find that even for weakly coupled dimers
the spectrum is very rich, consisting of many bound states.
The number of bound states depends on both
geometry of coupling and frustration. Many of the bound states can only
be seen by going to sufficiently high orders in the perturbation theory,
showing the extended character of the pair-attraction.
\end{abstract}

\pacs{}

]

\narrowtext

The study of bound states, life times and spectral weights for
multiparticle excitations remains a challenging problem
in many-body physics. From an experimental point of view, a variety
of probes on several low-dimensional magnetic and strongly
correlated electronic systems show spectral features associated
with multiparticle continuum and bound states. These features
in two-magnon Raman spectra, optical absorption spectra,
photoemission and even in neutron scattering spectra remain poorly
understood.
From a theoretical point of view, one of the most intriguing issues
is the role the growing number of bound states play in the 
confinement-deconfinement transition in spin-Peierls systems
as the spectrum completely changes from soliton-antisoliton
continuum to triplets, their bound states and continuum
\cite{Affleck:97}.
Existing computational approaches to these problems are not adequate.

In this letter, we present significant advances
in calculating multi-particle spectral properties
from high-order strong coupling expansions. At the heart
of our method is a generalization of Gelfand's linked cluster
expansion for single-particle excited states\cite{gelfand2} to 
multi-particle states. From a technical point
of view, our most notable achievement is the development
of an orthogonality transformation which leads to a
linked cluster theorem for multi-particle states even
with quantum numbers identical to the ground state.
Our method is quite distinct from the flow equation method of Wegner
\cite{Wegner:94}, which has been used recently by Knetter et al. 
\cite{Knetter:00} for also studying multi-particle spectral properties,
but
for a more restricted class of models with equispaced unperturbed
eigenvalues and an exact direct-product ground state.

We apply the method to systems of coupled spin-half dimers
in various one and two-dimensional geometries. The calculations greatly
simplify when we deal with models where the ground state is known
exactly to be a product of dimers, such as the
Shastry-Sutherland models in one and two dimensions
\cite{Shastry:81}. Here, we focus attention on the one-dimensional
systems. Results are presented for a spin-ladder system, 
the alternating spin-chain model, and the Shastry-Sutherland model.

Except for the spin-ladder model, the high order calculations produce
qualitatively unexpected results. There are multiple bound states with
$S=0$,
$1$ and $2$ even in the weakly-coupled dimer limit. The number of bound
states varies with frustration and many of them only show up when the
expansions are done to sufficiently high order. This presumably is due to
the extended range of the attractive interaction needed to see
the multiple bound states. The interval of k-values
where the bound states exist also changes with the expansion parameter.

We begin describing our method with a Hamiltonian
\begin{equation}
               H = H_{0} + \lambda H_{1},
\end{equation}
where the unperturbed part $H_{0}$ is exactly solvable, and
$\lambda$ is the perturbation parameter. The aim is to calculate
perturbation series in $\lambda$ for the eigenvalues of $H$ and other
quantities of interest. As is well known \cite{he,gelfand1}, the ground
state
energy and correlation functions have a `cluster addition property' and
hence 
can be calculated by linked cluster expansion. 

We wish to consider the 
excited-state many-particle sectors of the Hilbert space,
where a ``particle" may refer to
a lattice fermion, a spin-flip, or other excitation, depending on the
model at hand. 
        The key step is to `block diagonalize' the Hamiltonian on any
finite cluster to form an effective Hamiltonian, via 
an orthogonal transformation (here we will only consider real
Hamiltonians):
\begin{equation}
H^{\rm eff} = O^{T} H O 
\end{equation}
where
$O = e^{S}$
and S is real, antisymmetric.
This transformation is constructed order-by-order in perturbation
theory, so that the ground state sits in a block by itself, the
1-particle states (which form a degenerate manifold under $H_{0}$, in
general) form another block, the 2-particle states another block, and
so on. 
The off-diagonal blocks of S are determined by the requirement that the
off-diagonal blocks of $H^{\rm eff}$ vanish; and we choose the diagonal
blocks of S to be zero.

Let the matrix element of $H^{\rm eff}$ between initial 1-particle state 
$|{\bf i} \rangle$ and final 1-particle state $|{\bf j}\rangle$,
labelled according to their positions on the lattice, be
\begin{equation}
E_{1}({\bf i,j}) = \langle {\bf j}| H^{\rm eff}|{\bf
i}\rangle.
\end{equation}
This excited state
energy is not extensive, and does not obey the `cluster addition
property'. However, as shown by Gelfand\cite{gelfand2},
the ``irreducible" 1-particle matrix element
\begin{equation}
\Delta_{1}({\bf i,j}) = E_{1}({\bf i,j}) - E_{0} \delta_{{\bf i,j}}
\end{equation}
does have the `cluster addition property'.
Furthermore, for a translationally invariant
system, the 1-particle states are
eigenstates of momentum:
\begin{equation}
| {\bf k} \rangle = \frac{1}{\sqrt{N}}\sum_{{\bf j}}
\exp(i{\bf k} \cdot {\bf j} ) | {\bf j} \rangle 
\end{equation}
(where N is the number of sites in the lattice), with energy 
above the ground state of
\begin{equation}
\omega_{1}({\bf k}) = \sum_{{\bf \delta}} \Delta_{1}({\bf
\delta})\cos({\bf k} \cdot \delta) .
\end{equation}

\begin{figure}[ht]
 \begin{center} 
 \epsfig{file=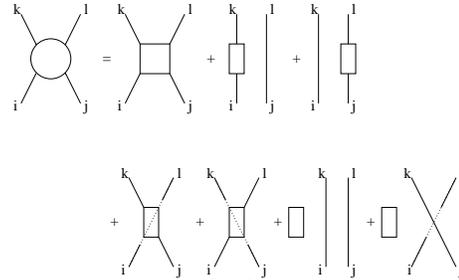, width=6cm}
 \caption[]
      {Decomposition of a 2-particle matrix element into irreducible 
      components. The round box denotes the full matrix element, the square 
      boxes the irreducible matrix elements, and the single line
denotes the delta function.}
\label{Decomposition} 
\end{center}
\end{figure}

To generalize to two-particle states, let
\begin{equation}
E_{2}({\bf i,j;k,l}) = \langle {\bf k,l} | H^{\rm eff} | {\bf i,j}
\rangle
\end{equation}
be the matrix element between initial 2-particle state $ | {\bf i,j}
\rangle$ and final state $ | {\bf k,l} \rangle$. To obtain a
quantity obeying the cluster addition property, we must subtract the
ground-state energy and 1-particle contributions, to form the
irreducible 2-particle matrix element (see Fig. \ref{Decomposition}):
\begin{eqnarray}
\Delta_{2}({\bf i,j;k,l}) &  = & E_{2}({\bf i,j;k,l})
-E_{0}(\delta_{{\bf
i,k}}\delta_{{\bf j,l}} + \delta_{{\bf i,l}}\delta_{{\bf j,k}})
\nonumber\\
& & -\Delta_{1}({\bf i,k})\delta_{{\bf j,l}} - \Delta_{1}({\bf i,l})
\delta_{{\bf j,k}} \nonumber\\
& & - \Delta_{1}({\bf j,k}) \delta_{{\bf i,l}} -
\Delta_{1}({\bf j,l}) \delta_{{\bf i,k}}
\end{eqnarray}
This quantity is easily found to be {\it zero} for any cluster
unless {\bf i, j, k} and {\bf l} are all included in that cluster, and
it obeys the cluster addition property. The block
diagonalization ensures that two particles cannot ``annihilate" from one
cluster and ``reappear" on another disconnected one. Thus the matrix
elements of $\Delta_{2}$ can be expanded in terms of connected clusters
alone, which are rooted or connected to all four positions {\bf i, j, k,
l}.

Once the effective two-particle Hamiltonian is known,
we still have to solve the
Schr\"odinger equation. 
Consider for simplicity a 2-particle state of identical particles, or
one symmetric under particle exchange.
Then expand the 2-particle eigenstate
\begin{equation}
| \psi \rangle = \sum_{{\bf i > j}} f_{{\bf ij}} | {\bf i,j} \rangle,
\hspace{5mm} f_{{\bf ij}} = f_{{\bf ji}} ,
\end{equation}
the Schr{\" o}dinger equation takes the form
\begin{eqnarray}
(E -E_{0})f_{{\bf ij}}& -& \sum_{\bf k}[ \Delta_{1}({\bf
k,i})f_{{\bf kj}} +  \Delta_{1}({\bf k,j})f_{{\bf
ik}}] \nonumber\\
&=& \frac{1}{2} \sum_{{\bf k,l}} \Delta_{2}({\bf k,l;i,j})f_{{\bf kl}}
\nonumber\\
 & & - \Delta_{1}({\bf j,i})f_{{\bf jj}} - \Delta_{1}({\bf i,j})f_{{\bf
ii}}
,
\hspace{2mm} all \hspace{1mm}{\bf i,j}.
\label{20}
\end{eqnarray}
The fictitious amplitudes $f_{{\bf ii}}$ are 
{\it defined} by these
equations\cite{mattis}, and are introduced to simplify the Fourier
transform. 
Defining the centre-of-mass momentum $\bf K$ and the relative momentum
$\bf q$,
this can be turned into an integral equation:
\begin{eqnarray}
&&[E - E_{0} -2\sum_{{\bf \delta}} \Delta_{1}({\bf \delta})\cos({\bf
K \cdot \delta}/2)\cos({\bf q \cdot \delta})]f({\bf K,q}) = \nonumber\\
&&    \frac{1}{N} \sum_{{\bf
q'}}f({\bf K,q'})[\frac{1}{2}\sum_{{\bf
r,\delta_{1},\delta_{2}}}\Delta_{2}({\bf
r,\delta_{1},\delta_{2}})
 \cos({\bf K \cdot r})\cos({\bf q \cdot \delta_{1}}) \nonumber\\ 
&& \times \cos({\bf 
q' \cdot \delta_{2}})
- 2\sum_{{\bf \delta}}\Delta_{1}({\bf \delta})\cos({\bf K \cdot
\delta}/2)\cos({\bf q \cdot \delta})],
\end{eqnarray}
which can be solved by standard numerical
techniques like discretization. The spectrum of the discretized
Hamiltonian can be obtained by diagonalization. In this way the
spectrum and even the two particle density of states can be obtained.
The continuum is limited by the the maximum (minimum) of the energy of
two single particle excitations whose combined momentum is the center
of mass momentum, which serves as an independent check. 


We have used this method to investigate the low-energy excitation
spectrum of 
the 2-leg spin-$\frac{1}{2}$ Heisenberg ladder
\begin{equation}
H = \sum_i {[J {\bf S}_i\cdot {\bf S}_{i+1} 
        + J {\bf S}_i^{\prime}\cdot {\bf S}^{\prime}_{i+1}
        + J_{\perp} {\bf S}_i\cdot {\bf S}^{\prime}_{i}]} \;,
\label{H:ladder}
\end{equation}
where the interactions along the ladder ($J$) and along the
rungs ($J_{\perp}$) are assumed to be antiferromagnetic.

For $J/J_{\perp}<\infty$, the ground state of this model 
evolves smoothly from a product of
singlet states along the rungs of the ladder and has a
gapped excitation spectrum
\cite{Gopalan:94,Oitmaa:96,Eder:97}.  The occurrence of two-particle
bound states in this model has been shown by first-order
strong-coupling expansions \cite{Damle:98,Jurecka:99} as well as a
leading order calculation using the analytic Brueckner
approach\cite{Kotov:98,Kotov:99}.

\begin{figure}
 \begin{center} 
 \vskip -0.5cm
 \epsfig{file=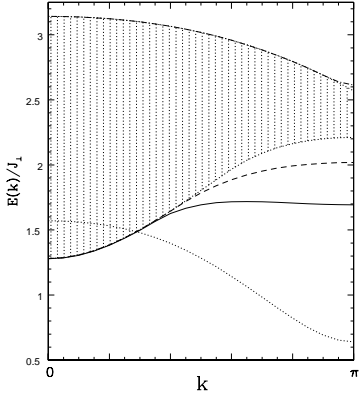, height=7cm}
  \vskip -1.2cm
 \caption[]
         {Low-energy excitation spectrum of the Heisenberg spin ladder for
          $J/J_{\perp} = 1/2$. Beside the two-particle continuum (gray 
          shaded) and the elementary triplet excitation (dotted line) there are
          three massive quasiparticles: a singlet bound state (solid line), 
          a triplet bound state (dashed line) and a
          quintet antibound state (dashed-dotted line).}
 \label{Ladder} 
 \end{center}   
\end{figure}

Starting from the dimerized ground state we have calculated series in 
 $J/J_{\perp}$ for $\Delta_2$  up to order 7 for singlet states, and
to order 12 for triplet and quintet states\cite{wwwsite}. Fig.
\ref{Ladder}
shows the generic shape of the two-particle continuum as well as the
low-lying
massive excitations. Beside the elementary triplet excitation the
spectrum
shows additional singlet ($S=0$) and triplet ($S=1$) excitations which
are
bound states of two elementary triplets.
In the vicinity of $K \approx\pi$ there is also an $S=2$ antibound state
above 
the continuum. 
At $J/J_{\perp} = 1/2$, we find the binding energy for the singlet bound
state at $K=\pi$ is $E_b/J_{\perp}=0.51$, substantially larger than 
the value 0.35 obtained in \cite{Kotov:99}.


Further, we have studied the occurrence of bound states in the alternating
Heisenberg chain
\begin{equation}
H = J \sum_i {[(1+(-1)^i\delta) {\bf S}_i\cdot {\bf S}_{i+1}
               + \alpha {\bf S}_i\cdot {\bf S}_{i+2}]} \;,
\label{H:chain}
\end{equation}
where the ${\bf S}_i$ are again spin-$\frac{1}{2}$ operators at site $i$,
$\alpha$ parameterizes a next-nearest neighbor coupling and $\delta$ is
the 
alternating dimerization.

\begin{figure}
 \begin{center} 
 \vskip -0.5cm
 \epsfig{file=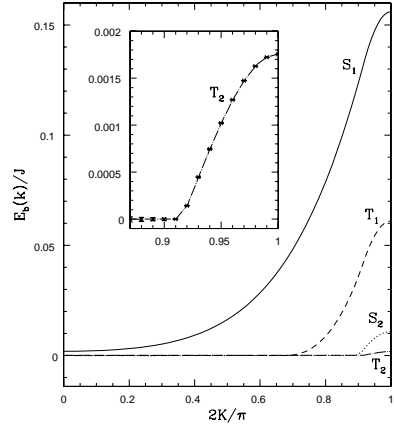, height=7.5cm}
 \vskip -1.2cm
 \caption[]
         {The binding energy $E_b$ for two singlet ($S_1$ and $S_2$) 
         and two triplet ($T_1$ and $T_2$) bound states
           versus momentum $K$ for the $J_1-J_2-\delta$ chain with
           $\delta=0.6$ and $\alpha=0$.
           The inset enlarges the region near $K=\pi/2$.}
 \label{Binding_K_0} 
 \end{center}
\end{figure}

\begin{figure}
 \begin{center} 
 \vskip -0.5cm
 \epsfig{file=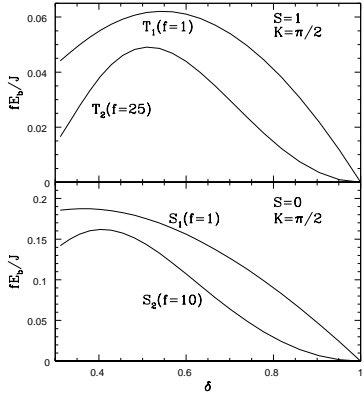, height=7cm}
 \vskip -1.2cm
 \caption[]
          {The scaled binding energy $f E_b$ at $K=\pi/2$ versus dimerization 
          $\delta$ for two singlet ($S_1$ and $S_2$) and two triplet ($T_1$ and $T_2$)
         states of the $J_1-J_2-\delta$ chain with $\alpha=0$, where $f$ 
          is a scale factor.}
 \label{Binding_delta} 
 \end{center}
\end{figure}

The two-particle excitations have been discussed in a leading order 
Brueckner ansatz calculation\cite{Bouzerar:98,Kotov:99b}, a second order
series expansion\cite{Barnes:99} and an RPA study\cite{Uhrig:96}.
With our new technique, we perform high-order series expansions in powers
of  $\lambda\equiv (1-\delta)/(1+\delta)$. 
Here we will only concentrate on the expansions for the following two
cases:

(1) $\alpha=$0, that is, without the second neighbor interaction. 
The series for $\Delta_2$ has been computed up to order 6 for singlet
bound 
states, and to order 11 for triplet and quintet states. 
Here we find two singlet, $S_1$ and $S_2$, and two triplet, $T_1$ and
$T_2$, 
bound states below the two-particle continuum.  
The binding energy of these bound states versus momentum $K$ is given in
Fig. \ref{Binding_K_0} for a rather large dimerization $\delta=0.6$.
The singlet $S_1$ exists for the whole range of momenta, while the 
triplet $T_1$ exists only in a limited range of momenta and the singlet
$S_2$ 
and  triplet $T_2$ bound states occur for momenta in the vicinity of the
band
maximum at $K=\pi/2$. 
The existence of the second pair of bound states has not been reported
by previous calculations, most likely due to a limited precision or a
general
incapability to deal with multiple bound states.
The binding energy at  $K=\pi/2$ versus dimerization $\delta$ is plotted
in 
Fig. \ref{Binding_delta}.
In the limit $\lambda\to 0$, the binding energies for $S_1$ and $T_1$ are
proportional to $\lambda$, as expected, since the formation of these
bound 
states is due to the attraction  of two triplets on neighboring sites. 
For $S_2$ and $T_2$, we find their binding energies are
proportional to
 $\lambda^2$.
This means that there is a strong enough effective attraction between
two triplets separated by a singlet dimer to form those bound states.

\begin{figure}[h]
 \begin{center} 
 \vskip -0.5cm
 \epsfig{file=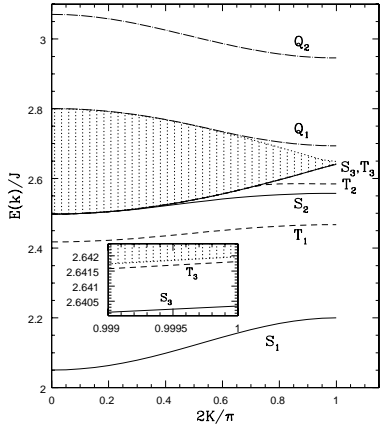, height=7.5cm}
 \vskip -1.2cm
 \caption[]
         {The excitation spectrum of the $J_1-J_2-\delta$ chain for
          $\delta=0.4$ and $\alpha=(1-\delta)/2$. Beside the two-particle
          continuum (gray shaded), there are
          three singlet bound states ($S_1$, $S_2$ and $S_3$), 
          three triplet bound states ($T_1$, $T_2$ and $T_3$) and two
          quintet antibound states ($Q_1$ and $Q_2$).
          The inset enlarges the region near 
          $K=\pi/2$ so we can see $S_3$ and $T_3$ below the continuum.}
 \label{mk_p5} 
 \end{center}
\end{figure}

(2) $\alpha = (1- \delta)/2$, that is, the expansion is along the
disorder 
line where the ground states are known exactly. The series for $\Delta_2$
has 
been computed up to order $\lambda^{19}$ for 2-particle singlet, triplet
and 
quintet states.
The two-particle excitation spectrum for $\delta=0.4$ is shown 
in Fig. \ref{mk_p5}. 
Here we find that there are three singlet and three triplet bound states
below
the two-particle continuum, and two quintet antibound states above the 
continuum. The energy gap at $K=\pi/2$ for one of the singlet bound
states, 
 $S_1$, is $1+3 \delta $ exactly.
In the limit $\lambda\to 0$, the binding energies for $S_n$ and $T_n$
($n=1,2$)
are of order $\lambda^n$, just as for $\alpha=0$, but that for $S_3$
and $T_3$ are at most of order $\lambda^4$. This calculation demonstrates
the power of the method in bringing out the complex character of the 
pair-attraction in the frustrated model.

In conclusion, we have demonstrated a new powerful approach to calculate 
high-order series expansion for quantum Hamiltonian lattice models. 
The application to the Heisenberg spin ladder and the alternating
Heisenberg 
chain has resulted in precise calculations of the low-lying excitation
spectra 
of these models.
For the alternating Heisenberg chain it was shown that there are multiple
massive singlet and triplet excitations below the continuum, which depend
on
frustration.


This work was initiated at the Quantum Magnetism program at the ITP at 
UC Santa Barbara which is supported by US National Science Foundation 
grant PHY94-07194.
ST gratefully acknowledges support by the German National Merit
Foundation.
HM wishes to thank the Kyoto Institute for Theoretical Physics for
hospitality.
The work of ZW and CJH was supported by a grant from the Australian
Research 
Council: they thank the New South Wales Centre for Parallel Computing for
facilities and assistance with the calculations. RRPS is supported in
part
by NSF grant number DMR-9986948.

\end{document}